\RequirePackage{fix-cm}
\documentclass[smallextended]{svjour3}       % onecolumn (second format)
\smartqed  % flush right qed marks, e.g. at end of proof
\usepackage{graphicx}
\usepackage{amssymb}
\usepackage{amsmath}
\usepackage{color,pstcol}
\newcommand{\ymodifartitionehundredfifteenvthreestepone}[1]
{#1}
\newcommand{\ymodifartitionehundredfifteenvthreesteptwo}[1]
{#1}
\newcommand{\ymodifartitionehundredfifteenvthreestepthree}[1]
{#1}
\journalname{}
\begin{document}

%YDsourcedeb inclusion artiti115v1step15
% Rq : j'ai verifie qu'il reclame \titlerunning
\title{$N$-qubit system in a pure state: a necessary
and sufficient condition for unentanglement}
\titlerunning{$N$-qubit system in a pure state: a
condition for unentanglement}
%YDvariante artiti115v1step15:
% \titlerunning{A condition for unentangled pure states of an $N$-qubit system}
%
%YDsourcefin inclusion artiti115v1step15
%YDsupprime artiti115v1step15 (remplace par ci-dessus):
% """
% \title{Insert your title here%\thanks{Grants or other notes
% %about the article that should go on the front page should be
% %placed here. General acknowledgments should be placed at the end of the article.}
% }
% \subtitle{Do you have a subtitle?\\ If so, write it here}
% 
% %\titlerunning{Short form of title}        % if too long for running head
% """

%YDsourcedeb inclusion artiti115v1step15
\author{Alain Deville
\and
Yannick Deville}
%YDsourcefin inclusion artiti115v1step15
%YDsupprime artiti115v1step15 (remplace par ci-dessus):
% """
% \author{First Author         \and
%         Second Author %etc.
% }
% """


%YDsourcedeb inclusion artiti115v1step15
\institute{Alain Deville \at
IM2NP UMR 7334, Aix-Marseille Universit\'{e}, CNRS, F-13397 Marseille,
France\\
\email{alain.deville@univ-amu.fr}\\
ORCID: 0000-0001-5246-8391
%YDsupprime artiti115v1step15:
% """
% \\
% Tel.: +33 5 61 33 28 24\\
% Fax: +33 5 61 33 28 40\\
% \email{alain.deville@univ-amu.fr}
% """           
\and
Yannick Deville \at
IRAP (Institut de Recherche en Astrophysique et Plan\'{e}tologie),
Universit\'{e} de
Toulouse, UPS, CNRS, CNES, 14 avenue Edouard Belin, F-31400 Toulouse, France\\
\email{yannick.deville@irap.omp.eu}\\
ORCID: 0000-0002-8769-2446
}
%YDsourcefin inclusion artiti115v1step15
%YDsupprime artiti115v1step15 (remplace par ci-dessus):
% """
% \institute{F. Author \at
%               first address \\
%               Tel.: +123-45-678910\\
%               Fax: +123-45-678910\\
%               \email{fauthor@example.com}           %  \\
% %             \emph{Present address:} of F. Author  %  if needed
%            \and
%            S. Author \at
%               second address
% }
% """

\date{Received: date / Accepted: date}

\maketitle

\begin{abstract}
%YDsourcedeb inclusion artiti115v1step15
If a pure state of a qubit pair is developed over the four 
%YDsourcedeb inclusion artiti115v1step18
basis states,
%YDsourcefin inclusion artiti115v1step18
%YDsupprime artiti115v1step18 (remplace par ci-dessus):
% """
% states of the
% standard basis, 
% """
%
%
%
%YDsourcedeb inclusion artiti115v3
\ymodifartitionehundredfifteenvthreestepthree{an equality
between the four coefficients of that development, verified if and only if
that state is unentangled, is already known.}
%YDsourcefin inclusion artiti115v3
%YDsupprime artiti115v3 (remplace par ci-dessus):
% """
% it is known that an equality between the four coefficients
% of that development exists if and only if that state is 
% %YDsourcedeb inclusion artiti115v1step15
% unentangled. 
% """
%
%
%
T%
%YDsourcefin inclusion artiti115v1step15
%YDsupprime artiti115v1step15 (remplace par ci-dessus):
% """
% unentangled\textbf{.
% T}
% """
his paper considers an arbitrary pure state of an $N$-qubit system,
developed over the $2^{N}$ 
%YDsourcedeb inclusion artiti115v1step18
basis states.
%YDsourcefin inclusion artiti115v1step18
%YDsupprime artiti115v1step18 (remplace par ci-dessus):
% """
% states of the generalized standard basis.\ 
% """
It is
shown that the state is unentangled if and only if a well-chosen collection
of\ $[2^{N}-(N+1)]$\ equalities between the $2^{N}$ coefficients of that
development is verified. The number of these equalities is large a soon as $%
N\gtrsim 10$, but it is shown that this set of equalities may be classified
into $(N-1)$ subsets, which should facilitate their manipulation. This
result should be useful e.g. in the contexts of Blind Quantum Source
Separation (BQSS)~and Blind Quantum Process Tomography (BQPT), with an aim
which should not be confused with that found when using the concept of
equivalence of pure states through local unitary transformations.%
%YDsourcefin inclusion artiti115v1step15
%YDsupprime artiti115v1step15 (remplace par ci-dessus):
% """
% Insert your abstract here. 
% """
%
%YDpasmis artiti115v1step15:
%YDsupprime artiti115v1step15:
% """
% Include keywords, PACS and mathematical
% subject classification numbers as needed.
% """
\end{abstract}
%
%
%
%YDsourcedeb artiti115v5: debut de partie copie (puis adaptee) de PourQIP2019IFFConditionV3SW_C.tex
% situe dans
% /home/ydeville/papier/ss/publi_ti2/artiti115_adeville_QIP_journal_critere_non_intrication/v5_egal_v3_soumise_a_qip_springer/step5_mail19_adeville_recu0718_v3_recu_v2_lu_et_reponse_a_reviewer
\section{Introduction \label{SectionIntroduction copy(1)}}

If both parts $S_{1}$ and $S_{2}$ of a bipartite quantum system $S$ are
initially prepared in a pure state, then, at a time scale allowing one to
neglect any coupling between $S$ and the rest of the universe, $S$ may be
described as being in a time-dependent pure state $\mid \Psi (t)>$ which
obeys the Schr\"{o}dinger equation.\ But, at that time scale, if an internal
coupling transiently exists between $S_{1}$ and $S_{2},$ then, after it
disappeared, $\mid \Psi (t)>$ often cannot be equated with a tensorial
product $\mid \Psi _{1}(t)>\otimes \mid \Psi _{2}(t)>$ describing $S_{1}$
and $S_{2}$ respectively: $\mid \Psi (t)>$ is entangled (or not separable) 
\cite{Schrodinger1935}, \cite{Buchleitner2009}. Entanglement plays a
significant role in Quantum Information (QI) \cite{Horodecki2009}, e.g. in
the context of Quantum Computing. The idea of a Quantum Computer (QC) dates
back to the 1980s\textbf{\ }\cite{Feynman1985}, \cite{Deutsch1985},\textbf{\ 
}and the word \textit{qubit} to name the basic cell of the Quantum Computer
appeared in 1995 (cf. \cite{Schumacher1995} and its acknowledgements). The
basic components of the future QC should be the qubit, the quantum register
- a quantum device consisting of several qubits- and the quantum gate - a
device aimed at controlling qubits and registers. It is presently possible
to claim that Quantum Computing research is coming of age \cite{Matsuura2019}%
, and qubits, registers and gates on one side, quantum algorithms using
abstract qubits on the other are under development.\ Qubits are generally
supposed to have been initially prepared in a pure state, which is not
always easily achieved with physical qubits, as shown e.g. for nuclear spins
in \cite{Nielsen2000}\textbf{\ (}cf.\ its page 324). Coupling between qubits
induces entanglement between initially unentangled qubit pure states. In the
context of QI, entanglement may be either desired, e.g. when considering a
QC, because it may allow some form of parallel computing (Deutsch speaks of
quantum parallelism \cite{Deutsch1985}), or avoided, because e.g. coupling
with the environment may cause decoherence, then stopping a calculation. It
should then be useful to be able to know whether a given pure state of a
system of abstract qubits is entangled or not.

The aim of this paper is to establish a necessary and sufficient
unentanglement condition (for breviety, an \textit{iff }condition) for a
system consisting of $N$ distinguishable qubits and which is in a pure
state. This iff condition should generalize a result which we recently
established for $N=2$ qubits \cite{DevilleA2017}, in the context of a
subfield of QI, based on Blind Source Separation (BSS), which started around
1985 in a classical context and which is now a mature field. In BSS \cite%
{Comon2010}, \cite{DevilleY2016}, typically, a set of users (the Writer)
presents a set of simultaneous signals (input signals, called sources) at
the input of a multi-user communication system (the Mixer). The sources,
compelled to possess some general properties (e.g. mutual statistical
independence), are then combined (mixed, in the BSS sense) in the Mixer.
Another set of users (the Reader) receives the signals arriving at the Mixer
output. The Writer possibly knows the sources, but the Reader does not know
them, and cannot access the inputs of the Mixer. That Mixer uses one or
several parameter values, unknown to the Reader, who only knows some of its
general properties. The Reader's final task is the restoration of the
sources (possibly up to some so-called acceptable indeterminacies) from the
signals at the Mixer output, during the \textquotedblleft inversion
phase\textquotedblright . An intermediate task is the determination of the
unknown parameters of the Mixer, or of its inverse, made during an
\textquotedblleft adaptation phase\textquotedblright . Since 2007, we have
been developing a quantum version of BSS, namely Blind Quantum Source
Separation (BQSS). In our previous papers (\cite{DevilleY2017} and
references therein\texttt{)}, we considered two distinguishable qubits
numbered 1 and 2, first prepared in a pure unentangled state $\mid \Psi
(t_{w})>$, by the Writer, at time $t_{w}$. The qubit pair is isolated from
the rest of the world between $t_{w}$ and the time $t_{r}$ when the Reader
can operate. At time $t_{r},$ the qubit pair is therefore still in a pure
state $\mid \Psi (t_{r})>$ but, because of an undesired coupling between the
qubits (viewed as the action of a Mixer), $\mid \Psi (t_{r})>$ is generally
\ entangled. The Reader's task is the restoration of $\mid \Psi (t_{w})>$ $=$
$\mid \Psi _{1}(t_{w})>\otimes \mid \Psi _{2}(t_{w})>$ from $\mid \Psi
(t_{r})>$ (again possibly up to some acceptable indeterminacies).\ In 2012,
in this journal, we presented a paper devoted to methods for separating
quantum sources \cite{DevilleY2012}. We recently introduced Blind Quantum
Process Tomography (BQPT \cite{DevilleY2015}, \cite{DevilleY2017IFAC}), the
blind version of Quantum Process Tomography (QPT \cite{Nielsen2000}), which
is defined in Section \ref{SectionDiscussion}. The iff condition hereafter
established is first aimed at extending the solution of the BQSS\ and BQPT
problems beyond the qubit pair, but it is hoped that it could be used in
other contexts, e.g that of the QC.

In Section \ref{ExistingCriteria}, we first recall the iff condition for $%
N=2 $ qubits, and then present already existing criteria related to pure
states or statistical mixtures of quantum systems composed of several parts,
in situations which are more or less different from the present one -
qubits, in arbitrary number. In\ Section \ref{SectionTowardsaGeneralization}
the method to be followed for finding this iff condition is presented.\ This%
\textit{\ iff} condition is established in Section \ref{SectionFinding a
necessary and sufficient}, through an iterative process. The results and
some of their applications are discussed in Section \ref{SectionDiscussion}.
The possibility of using the von Neumann concept is briefly tackled in the
Appendix.

\section{About existing unentanglement criteria\label{ExistingCriteria}}

In our previous papers, qubits were supposed to be physically implemented as
spins 1/2. We hereafter first recall the notations used for writing an
arbitrary pure state $\mid \Psi >$ of a qubit pair.\texttt{\ }$\mid \Psi >$%
\texttt{\ }is developed as:%
\begin{eqnarray}
&\mid &\Psi 
%YDsourcedeb inclusion artiti115v5
>
=
%YDsourcefin inclusion artiti115v5
%YDsupprime artiti115v5 (remplace par ci-dessus):
% """
% \text{
% %YDsupprime artiti115v5:
% % \TEXTsymbol
% {>}=}
% """
\sum_{i=1}^{4}c_{i}\mid i>\text{,\texttt{\ 
}where}  \label{Definition_ci_pairequbits} \\
&\mid &1>=\mid ++>,\text{ }\mid 2>=\mid +->\text{, }\mid 3>=\mid -+>,\text{ }%
\mid 4>=\mid -->  \label{Definition_Etats_i}
\end{eqnarray}%
where $\mid ++>$ is an abbreviation for $\mid 1,+>\otimes \mid 2,+>,$ $\mid
1,+>$ being the eigenstate for the eigenvalue $1/2$ in the standard basis of
qubit 1.\ We now consider an arbitrary $N-$qubit system, and generalize this
writing: for qubit number $k$ in an $N$-qubit system:

\begin{equation}
s_{zk}\mid k,\pm >=\pm \frac{1}{2}\mid k,\pm >  \label{Definitionksms}
\end{equation}%
and for a pure state $\mid \Psi >$ of this $N$-qubit system:%
\begin{equation}
\mid \Psi >=\sum_{i=1}^{2^{N}}c_{i}\mid i>  \label{Definition_ciNqubits}
\end{equation}%
where $\mid 1>\ =$ $\mid ++....++>$ (all qubits in state $\mid +>$),$~\mid
2^{N}>=\mid ---....-->$ (all qubits in state $\mid ->$).\ Qubit $1$ is in
state $\mid +>$ in the first $2^{N-1}$ states, and in state $\mid ->$ in the 
$2^{N-1}$ remaining states. $\mid 2^{N-1}+1>$ $=\mid -+..++>$ (all qubits in
state $\mid +>$, except qubit 1, in state $\mid ->$), $\mid 2^{N}-2>$ $=\mid
---....+->$ (all qubits in state $\mid ->$, except qubit $N-1$, in state $%
\mid +>$).

We immediately prove (for $c_{1}\neq 0)$ the following property, which we
established in more detail (including the $c_{1}=0$ case) in \cite%
{DevilleA2017}: a pure state $\mid \Psi >$ of a qubit pair is unentangled if
and only if the $c_{i}$ coefficients obey the equality:%
\begin{equation}
c_{1}c_{4}=c_{2}c_{3}.  \label{cns_NonIntrication2Qubits}
\end{equation}%
\texttt{\ }If $c_{1}\neq 0$, $\ \mid \Psi >$ can always be written as:%
\begin{equation}
\mid \Psi >=c_{1}(\mid ++>+\frac{c_{2}}{c_{1}}\mid +->+\frac{c_{3}}{c_{1}}%
\mid -+>+\frac{c_{4}}{c_{1}}\mid -->).
\end{equation}%
When $c_{1}c_{4}=c_{2}c_{3}$,~$\mid \Psi >$ can then be written as:%
\begin{eqnarray}
&\mid &\Psi >=c_{1}(\mid ++>+\frac{c_{2}}{c_{1}}\mid +->+\frac{c_{3}}{c_{1}}%
\mid -+>+\frac{c_{2}c_{3}}{c_{1}^{2}}\mid -->) \\
&=&c_{1}(\mid +>+\text{\ }\frac{c_{3}}{c_{1}}\mid ->)\otimes (\mid +>+\frac{%
c_{2}}{c_{1}}\mid ->)
\end{eqnarray}%
\bigskip which proves that when $N=2$ and $c_{1}c_{4}=c_{2}c_{3},$ $\mid
\Psi >$ is unentangled.

Conversely, if $\mid \Psi >$ is unentangled, it may be written as:%
\begin{equation}
\mid \Psi >=(a_{1}\mid +>+b_{1}\mid ->)\otimes (a_{2}\mid +>b_{2}\mid ->).
\label{Pour recip Pour N=2}
\end{equation}%
Eq. (\ref{Pour recip Pour N=2}) must be consistent with Eq. (\ref%
{Definition_ci_pairequbits}), which imposes the following relations:%
\begin{equation}
c_{1}=a_{1}a_{2}\text{, \ \ }c_{2}=a_{1}b_{2},\text{ \ \ }c_{3}=b_{1}a_{2},%
\text{ \ \ }c_{4}=b_{1}b_{2}
\end{equation}%
and therefore $c_{1}c_{4}=c_{2}c_{3}.$

In this paper, the proposed \textit{iff }condition will take the form of a
set of equalities between the $c_{i}$ coefficients. In Section \ref%
{SectionFinding a necessary and sufficient} it will e.g. be shown that $\mid
\Psi >$ is unentangled, when $N=3,$ if and only if the following two subsets
(S1, S2) of equalities, corresponding to\ a total of four independent
equalities, are simultaneously verified:%
\begin{eqnarray}
S_{1} &:&\text{ \ \ }\frac{c_{1}}{c_{2}}=\frac{c_{3}}{c_{4}}=\frac{c_{5}}{%
c_{6}}=\frac{c_{7}}{c_{8}}  \label{Subset1forN=3} \\
S_{2} &:&\ \ \ c_{2}c_{8}=c_{4}c_{6}  \label{Subset2forN=3}
.
\end{eqnarray}%
Coming now to existing criteria, one should first mention the Schmidt and
the Peres-Horodecki ones. The so-called Schmidt decomposition allows one to
express an \textit{iff }condition for pure states of a bipartite system, in
which the dimension of each part of the system is not restricted to $2.$\
And an extension of the concept of entanglement to statistical mixtures is
somewhat possible through the notions of separability and of entanglement
witnesses \cite{Buchleitner2009}. The Peres-Horodecki criterion \cite%
{Horodecki2009} is an \textit{iff }condition for the separability of the
density matrix describing a state of a bipartite system, valid when the
dimensions of the state spaces of $S_{1}$and $S_{2}$ are low. In \cite%
{DevilleA2017}, it was explained why these criteria are not appropriate for
the context of BQSS. They are both restricted to bipartite systems, and
therefore should presently be discarded, since this paper is devoted to the
general case $N>2.$\ With these\ presently strong restrictions, their aim is
not very different from the one in this paper.

On the contrary, in the context of quantum communications, it is usual to
speak of Local Unitary (LU) transformations of pure states, and when it is
spoken of equivalent states the aim is different from the one in e.g.\ BQSS.
If $\mid \Psi (1,2,...N)>$ is some pure state of a system composed of $N$
distinguishable particles, and $U$ a unitary operator acting on $\mid \Psi
(1,2,...N)>,$ then producing a transformed pure state $U\mid \Psi
(1,2,...N)>,$ the transformation is said to be local if $U=U(1)\otimes
U(2)...\otimes U(N)$, where $U(1),$\ $U(2),$ $...$ $U(N)$ are themselves
unitary operators, each acting upon a single particle. The reason for the
choice of the word \textit{Local }may be guessed if 1) one considers the
transformation of an unentangled state $\mid \Psi (1,2,...N)>$, 2a) one
considers two qubits, 1 in space-time zone 1, and 2 in space-time zone 2,
separated by a spacelike interval, 2b) one imagines two experimenters,
A(lice) and B(ob), A being able to access qubit 1 (only) and B qubit 2
(only), 2c) this situation is true for all distinct pairs of an $N$-qubit
system. On the contrary, an instance of the effect of a non-LU
transformation on an unentangled state is given in Section \ref%
{SectionDiscussion}. Once LU transformations have been defined, two pure
states are said to be equivalent if they differ by an LU transformation only
(cf.\ e.g. the review article \cite{Walter2017}).\ When trying to define a
degree of entanglement in order to classify all the possible pure states of
a multipartite system, one is led to make no distinction between equivalent
states. On the contrary, in the context of BQSS, this concept of equivalence
through an LU transformation plays no role.\ If a given pure state $\mid
\Psi >$ is written at the input of the Mixer, and a state $\mid \Phi >$ is
read at its output, the aim is not to get back some state $\mid \Xi >$
equivalent to $\mid \Psi >$, with the meaning that $\mid \Xi >$ and $\mid
\Psi >$ differ by an LU transformation, but to get back state $\mid \Psi >$
itself, possibly up to some acceptable (in fact, far weaker than an
arbitrary LU transformation) indeterminacies.

\section{Towards a generalization of the relation between the c$_{i}$
existing for a qubit pair\label{SectionTowardsaGeneralization}}

In order to extend entanglement-based BQSS methods beyond the simplest case,
the qubit\ pair $(N=2)$, it is highly desirable to find a set of relations,
if it does exist, which, when $N>2,$ could be substituted for the equality $%
c_{1}c_{4}=c_{2}c_{3}.$\ We have to find a collection of equalities between
the $c_{i}$ which are obeyed if and only if $\mid \Psi >$ is unentangled. We
will consider only normalized states: $<\Psi $ $\mid \Psi >=1.$ Moreover
only the projector $\mid \Psi >$ $<\Psi $ $\mid $ has a physical meaning:
one should not distinguish $\ $between $\mid \Psi >$ and $e^{i\eta }$ $\mid
\Psi >$ ($\eta $ is any real number).\ Therefore, if the complex numbers $%
c_{i}$ are written $c_{i}=\rho _{i}e^{i\varphi _{i}}$ ($\rho _{i}$ and $%
\varphi _{i}$ are real numbers and $i=1$, $2$,$...2^{N})$, then rather than
the $2^{N}$ phases, only e.g. the $(2^{N}-1)$ phase differences ($\varphi
_{i}-\varphi _{1})$ are meaningful when defining an arbitrary pure state $%
\mid \Psi >.$\ Consequently, an arbitrary pure state $\mid \Psi >$ of the $N$%
-qubit system is defined by the value of $(2^{N+1}-2)$ independent real
numbers: $(2^{N}-1)$ moduli and $(2^{N}-1)$ phases. However, an unentangled
pure state may be written as:%
\begin{equation}
\mid \Psi _{ue}>=\mid \psi _{1}>\otimes \mid \psi _{2}>\otimes ...\mid \psi
_{i}>\otimes ...\otimes \mid \psi _{N}>
\end{equation}%
where each ordered factor describes the state of a given qubit, and
therefore depends upon two real numbers (a modulus, a phase).\ An
unentangled state $\mid \Psi _{ue}>$ therefore depends upon $2N$ real
numbers only. Then, if $\mid \Psi _{ue}>$ is developed following (\ref%
{Definition_ciNqubits}), there should exist $2[2^{N}-(N+1)]$ relations
between the $2^{N+1}$ real quantities \{$\rho _{i},$ $\varphi _{i}$\}
(besides those expressing that only $\mid \Psi ><\Psi \mid $ has a physical
meaning and that $\mid \Psi >$ is normalized). This result strengthens the
hope that there may exist, between the $c_{i}$ coefficients, a set of $%
[2^{N}-(N+1)]$ equalities which are verified if and only if $\mid \Psi >$ is
unentangled.\ This is presently only a hope, since normalization of $\mid
\Psi >$ and the physical meaning of $\mid \Psi >$ $<\Psi $ $\mid $ lead to
two constraints between real numbers, not to a constraint between the $c_{i}$
themselves.\ The present paper will \textit{not }try to directly prove the
existence of such relations between the $c_{i}$ for unentangled and only
unentangled pure states, but will rather use an iterative approach in order
to try and establish such general relations.

\section{Finding a necessary and sufficient condition: an iterative approach 
\label{SectionFinding a necessary and sufficient}}

We will first examine the $N=3$ and $N=4$ cases in some detail, starting
with $N=3.$\ It is supposed that $c_{i}\neq 0$ for any $i$ value. The case
when $c_{i}=0$ for at least one $i$ value will be discussed in Section \ref%
{SectionDiscussion}.

When $N=3$, any pure state may be written as:%
\begin{eqnarray}
\mid \Psi >
&
=
&(\mid ++>+\frac{c_{3}}{c_{1}}\mid +->+\frac{c_{5}}{c_{1}}\mid
-+>+\frac{c_{7}}{c_{1}}\mid -->)\otimes c_{1}\mid +>  \notag \\
&
&
+( \mid ++>+\frac{c_{4}}{c_{2}}\mid +->+\frac{c_{6}}{c_{2}}\mid -+>+\frac{%
c_{8}}{c_{2}}\mid -->)\otimes c_{2}\mid ->  \label{PsiArbitraireN=3}
\end{eqnarray}%
where $\mid +>$ in the writing $c_{1}\mid +>$ and $\mid ->$ in the writing $%
c_{2}\mid ->$ refer to a state of qubit no. $3.$ When%
\begin{equation}
\frac{c_{3}}{c_{1}}=\frac{c_{4}}{c_{2}},\frac{c_{5}}{c_{1}}=\frac{c_{6}}{%
c_{2}},\frac{c_{7}}{c_{1}}=\frac{c_{8}}{c_{2}}  \label{PourN=3,fournitS1}
\end{equation}%
it is possible to write $\mid \Psi >$\ as:%
\begin{equation}
\mid \Psi >=\underset{\mid \Psi (1,2)>}{\underbrace{(\mid ++>+\frac{c_{4}}{%
c_{2}}\mid +->+\frac{c_{6}}{c_{2}}\mid -+>+\frac{c_{8}}{c_{2}}\mid -->)}}%
\otimes \underset{\mid \Psi (3)>}{\underbrace{(c_{1}\mid +>+c_{2}\mid ->)}}.
\end{equation}%
When moreover the coefficients of $\mid \Psi (1,2)>$ obey Eq. (\ref%
{cns_NonIntrication2Qubits}), which is presently written as:%
\begin{equation}
\frac{c_{8}}{c_{2}}=\frac{c_{4}}{c_{2}}.\frac{c_{6}}{c_{2}},\text{ }i.e.%
\text{ \ }c_{2}c_{8}=c_{4}c_{6}  \label{Subset2N=3}
\end{equation}%
\bigskip then $\mid \Psi (1,2)>$ is an unentangled state, and $\mid \Psi >$\
is unentangled.\ Therefore, when $N=3,$\ if subsets (\ref{Subset1forN=3})
and (\ref{Subset2forN=3}) are obeyed, $\mid \Psi >$\ is unentangled (subset (%
\ref{Subset1forN=3}) is equivalent to the three equations (\ref%
{PourN=3,fournitS1})).

Conversely, if $\mid \Psi >$\ is unentangled, it can be written as:%
\begin{equation}
\mid \Psi (1,2,3)>=\underset{\text{unentangled}}{\underbrace{%
(\sum_{i=1}^{4}C_{i}\mid i>)}}\otimes (a_{3}\mid +>+b_{3}\mid ->)
\label{Psi(123)NonIntrique}
\end{equation}%
where the $C_{i}$ coefficients obey the relation $C_{1}C_{4}$ $=$ $C_{2}C_{3}
$. Identifying Eq. (\ref{Psi(123)NonIntrique}) with Eq. (\ref%
{Definition_ciNqubits}) (with $N=3$), one gets:%
\begin{equation}
c_{1}=C_{1}a_{3},\text{ }c_{2}=C_{1}b_{3},\ c_{3}=C_{2}a_{3},\text{ }...%
\text{ }c_{8}=C_{4}b_{3}  \label{LesCiPourN=3}
\end{equation}%
and therefore Eqs. (\ref{Subset1forN=3}) and (\ref{Subset2forN=3}) (i.e.
Conditions $S_{1}$ and\ $S_{2}$) are obeyed.

When $N=3$, $|\Psi >$ is therefore unentangled if and only if the two
subsets of equalities (\ref{Subset1forN=3}) and (\ref{Subset2forN=3}),
expressing a total of $4$ (independent) equalities, are obeyed. Therefore,
the above-expressed hope of finding an \textit{iff }condition for
unentanglement using $[2^{N}-(N+1)]$ relations between the $c_{i}$ is
satisfied when $N=3.$

Similarly, when $N=4$, any pure state may be written as:%
\begin{eqnarray}
\mid \Psi >
&
=
&
(\mid +++>+\frac{c_{3}}{c_{1}}%
\mid ++->+...+\frac{c_{15}}{c_{1}}\mid --->)\otimes c_{1}\mid +>  \notag \\
&
&
+( \mid +++>+\frac{c_{4}}{c_{2}}%
\mid ++->+...+\frac{c_{16}}{c_{2}}\mid --->)\otimes c_{2}\mid ->
\label{PsiArbitraireN=4}
\end{eqnarray}%
where now $\mid +>$ in the writing $c_{1}\mid +>$ and $\mid ->$ in the
writing $c_{2}\mid ->$ refer to a state of qubit no. $4$.\ When the
following set of equalities is satisfied:%
\begin{equation}
\frac{c_{3}}{c_{1}}=\frac{c_{4}}{c_{2}},\frac{c_{5}}{c_{1}}=\frac{c_{6}}{%
c_{2}},...,\frac{c_{15}}{c_{1}}=\frac{c_{16}}{c_{2}},
\end{equation}%
\bigskip which can also be written as:%
\begin{equation}
\frac{c_{1}}{c_{2}}=\frac{c_{3}}{c_{4}}=\frac{c_{5}}{c_{6}}=...=\frac{c_{15}%
}{c_{16}},  \label{Subset1N=4}
\end{equation}%
\bigskip $\mid \Psi >$ may be written as:%
\begin{equation}
\mid \Psi >=\underset{\mid \Psi (1,2,3)>}{\underbrace{(\mid +++>+\frac{c_{4}%
}{c_{2}}\mid ++->+...\frac{c_{16}}{c_{2}}\mid --->)}}\otimes (c_{1}\mid
+>+c_{2}\mid ->).
\end{equation}%
When moreover the coefficients of $\mid \Psi (1,2,3)>$ obey subsets (\ref%
{Subset1forN=3}) and (\ref{Subset2forN=3}), presently written as:%
\begin{eqnarray}
\frac{c_{2}}{c_{4}} &=&\frac{c_{6}}{c_{8}}=\frac{c_{10}}{c_{12}}=\frac{c_{14}%
}{c_{16}}  \label{Subset2N=4} \\
c_{4}c_{16} &=&c_{8}c_{12},  \label{Subset3N=4}
\end{eqnarray}%
$\bigskip \mid \Psi >$ is unentangled. Therefore, when $N=4,~$if the three
subsets of Eqs. (\ref{Subset1N=4}), (\ref{Subset2N=4}) and (\ref{Subset3N=4}%
) are verified, $\mid \Psi >$ is unentangled.

Conversely, if $\mid \Psi >$\ is unentangled, it can be written as:%
\begin{equation}
\mid \Psi (1,2,3,4)>=\underset{\text{unentangled}}{\underbrace{%
(\sum_{i=1}^{8}C_{i}\mid i>)}}\otimes (a_{4}\mid +>+b_{4}\mid ->)
\label{Psi(1234NonIntrique}
\end{equation}%
where the $C_{i}$ coefficients now obey the following sets of relations:%
\begin{eqnarray}
\frac{C_{1}}{C_{2}} &=&\frac{C_{3}}{C_{4}}=\frac{C_{5}}{C_{6}}=\frac{C_{7}}{%
C_{8}}  \label{N=4Psi(123)NonIntriqueS1} \\
C_{2}C_{8} &=&C_{4}C_{6}  
.
\label{N=4Psi(123)NonIntriqueS2}
\end{eqnarray}%
Now identifying Eq. (\ref{Psi(1234NonIntrique}) with Eq. (\ref%
{Definition_ciNqubits}) (with $N=4$), one gets:%
\begin{equation}
c_{1}=C_{1}a_{4},\text{ }c_{2}=C_{1}b_{4},\text{\ }%
c_{3}=C_{2}a_{4}.....c_{16}=C_{8}b_{4}.  \label{LesCiPourN=4}
\end{equation}%
\ These equalities were obtained through the same process as in Eq. (\ref%
{LesCiPourN=3}), and therefore lead to the same type of equalities as in Eq.
(\ref{Subset1forN=3}), but with now sixteen and not eight $c_{i}$\
coefficients, namely:%
\begin{equation}
\frac{c_{1}}{c_{2}}=\frac{c_{3}}{c_{4}}=...=\frac{c_{15}}{c_{16}}.
\end{equation}%
Moreover, from Eqs. (\ref{N=4Psi(123)NonIntriqueS1}), (\ref%
{N=4Psi(123)NonIntriqueS2}) and (\ref{LesCiPourN=4}), one gets:%
\begin{eqnarray}
\frac{c_{2}}{c_{4}} &=&\frac{c_{6}}{c_{8}}=\frac{c_{10}}{c_{12}}=\frac{c_{14}%
}{c_{16}} \\
c_{4}c_{16} &=&c_{8}c_{12}.
\end{eqnarray}%
Therefore, when $N=4,$ $\mid \Psi >$ is unentangled if and only if the
following three subsets of equalities are verified:%
\begin{eqnarray}
S_{1} &:&\ \ \ \ \frac{c_{1}}{c_{2}}=\frac{c_{3}}{c_{4}}=...=\frac{c_{15}}{%
c_{16}}  \label{Subset1forN=4} \\
S_{2} &:&\ \ \ \ \frac{c_{2}}{c_{4}}=\frac{c_{6}}{c_{8}}=\frac{c_{10}}{c_{12}%
}=\frac{c_{14}}{c_{16}}  \label{Subset2forN=4} \\
S_{3} &:&\ \ \ \ c_{4}c_{16}=c_{8}c_{12}.  \label{Subset3forN=4}
\end{eqnarray}%
When $N=4$, this \textit{iff }condition is therefore expressed through 3
subsets of equations, expressing a total of $11$ equalities, which is also
again the value of $[2^{N}-(N+1)]$, now for $N=4.$

Now taking an arbitrary $N$ value, one may write any pure state $\mid \Psi >$
as:%
\begin{eqnarray}
\mid \Psi >
&
=
&
(\mid ++...+>+\frac{c_{3}}{c_{1}}\mid
++...->+...+\frac{c_{2^{N}-1}}{c_{1}}\mid --...->)\otimes c_{1}\mid +> 
\notag \\
&
&
+(\mid ++...+>+\frac{c_{4}}{c_{2}}\mid
++..->+...+\frac{c_{2^{N}}}{c_{2}}\mid --...->)\otimes c_{2}\mid ->
\notag \\
\label{PsiArbitraireNArbitraire}
\end{eqnarray}%
which generalizes Eqs. (\ref{PsiArbitraireN=3}) and (\ref{PsiArbitraireN=4}%
). The reasoning which led to Eq. (\ref{Subset1forN=3}) and (\ref%
{Subset1forN=4}) now leads to:%
\begin{equation}
\frac{c_{1}}{c_{2}}=\frac{c_{3}}{c_{4}}=...=\frac{c_{2^{N}-3}}{c_{2^{N}-2}}=%
\frac{c_{2^{N}-1}}{c_{2^{N}}}.  \label{Subset1ArbitraryN}
\end{equation}%
This subset contains $(2^{N-1}-1)$ independent equalities. When they are
obeyed, $\mid \Psi >$ may be written as:%
\begin{equation}
\mid \Psi >=\mid \Psi (1,2,3,...N-1)>\otimes \mid \Psi (N)>.
\label{PsiproduitPsi(1.N-1)*Psi(N)}
\end{equation}%
This state $\mid \Psi >$ is unentangled if and only if, moreover, $\mid \Psi
(1,2,3,...N-1)>$ itself is unentangled. If the results obtained for $N=3$
and $N=4$ may be generalized, this is obtained when $(N-2)$ other subsets of
inequalities, corresponding to a total of $[2^{N}-(N+1)]$ $-$ $(2^{N-1}-1)$
equalities between the $c_{i},$ i.e. $(2^{N-1}-N)$ independent equalities,
are satisfied. This quantity is equal to $1$ if $N=3$ (subset $S_{2}$, Eq. (%
\ref{Subset2forN=3})), and to $4$ if $N=4$ (subsets $S_{2},$ Eq. (\ref%
{Subset2forN=4}), and $S_{3},$ Eq. (\ref{Subset3forN=4})).

We now momentarily suppose that the property established for $N=3$ and $N=4$
is true for $N-1$ (with $N-1\geq 4$), i.e. that $\mid \Psi >$ is unentangled
if and only if the following $(N-2)$\ subsets of equalities are
simultaneously verified:%
\begin{eqnarray}
S_{1} &:&\ \ \frac{c_{1}}{c_{2}}=\frac{c_{3}}{c_{4}}=...=\frac{c_{2^{N-1}-3}%
}{c_{2^{N-1}-2}}=\frac{c_{2^{N-1}-1}}{c_{2^{N-1}}}
\label{Subset1For(N-1)qubits} \\
S_{2} &:&\ \ \frac{c_{2}}{c_{4}}=\frac{c_{6}}{c_{8}}=...=\frac{c_{2^{N-1}-2}%
}{c_{2^{N-1}}} \\
S_{3} &:&\ \ \frac{c_{4}}{c_{8}}=\frac{c_{12}}{c_{16}}=...=\frac{%
c_{2^{N-1}-4}}{c_{2^{N-1}}} \\
&&.............  \label{Subset...For(N-1)qubits} \\
S_{N-3} &:&\ \ \frac{c_{2^{N-4}}}{c_{2\ast 2^{N-4}}}=\frac{c_{3\ast 2^{N-4}}%
}{c_{4\ast 2^{N-4}}}=\frac{c_{5\ast 2^{N-4}}}{c_{6\ast 2^{N-4}}}=\frac{%
c_{7\ast 2^{N-4}}}{c_{8\ast 2^{N-4}}} \\
S_{N-2} &:&\ \ c_{2^{N-3}}c_{4\ast 2^{N-3}}=c_{2\ast 2^{N-3}} c_{3\ast
2^{N-3}}.  \label{Subset(N-2)For(N-1)qubits}
\end{eqnarray}%
In order to help the reader anxious to see more explicitly the meaning of
the dots in Eq. (\ref{Subset...For(N-1)qubits}), and in Eq. (\ref%
{NquibtsSubset...}) hereafter, the 6 subsets for $N=7$ are all written at
the end of this section (four of them again with dots, but then with a
rather obvious meaning). We now establish that any ket $\mid \Psi >$ $%
=\sum_{i=1}^{2^{N}}c_{i}\mid i>$\ describing a pure state of an $N$-qubit
system is unentangled if and only if the $N-1$\ following subsets are all
obeyed: 
\begin{eqnarray}
S_{1} &:&\ \ \frac{c_{1}}{c_{2}}=\frac{c_{3}}{c_{4}}=...=\frac{c_{2^{N}-3}}{%
c_{2^{N}-2}}=\frac{c_{2^{N}-1}}{c_{2^{N}}}  \label{NquibtsSubset1} \\
S_{2} &:&\text{\ \ }\frac{c_{2}}{c_{4}}=\frac{c_{6}}{c_{8}}=...=\frac{%
c_{2^{N}-2}}{c_{2^{N}}}  \label{NqubitsSubset2} \\
S_{3} &:&\ \ \frac{c_{4}}{c_{8}}=\frac{c_{12}}{c_{16}}=...=\frac{c_{2^{N}-4}%
}{c_{2^{N}}} \\
&&.............  \label{NquibtsSubset...} \\
S_{N-2} &:&\ \ \frac{c_{2^{N-3}}}{c_{2\ast 2^{N-3}}}=\frac{c_{3\ast 2^{N-3}}%
}{c_{4\ast 2^{N-3}}}=\frac{c_{5\ast 2^{N-3}}}{c_{6\ast 2^{N-3}}}=\frac{%
c_{7\ast 2^{N-3}}}{c_{8\ast 2^{N-3}}} \\
S_{N-1} &:&\ \ c_{2^{N-2}} c_{4\ast 2^{N-2}}=c_{2\ast 2^{N-2}}c_{3\ast
2^{N-2}}.  \label{Subset(N-1)ForNqubits}
\end{eqnarray}

The approach already used for $N=3$ and $4$ is applied to an $N-$qubit
system. Any pure state may be written as:%
\begin{eqnarray}
\mid \Psi >
&
=
&
(\mid ++...+>+\frac{c_{3}}{c_{1}}\mid
++...->+...+\frac{c_{2^{N}-1}}{c_{1}}\mid --...->)\otimes c_{1}\mid +> 
\notag \\
&
&
+(\mid ++...+>+\frac{c_{4}}{c_{2}}\mid
++...->+...+\frac{c_{2^{N}}}{c_{2}}\mid --...->)\otimes c_{2}\mid ->
\notag \\
\end{eqnarray}%
where now $\mid +>$\ in the writing $c_{1}\mid +>$\ and $\mid ->$\ in the
writing $c_{2}\mid ->$\ refer to a state of qubit no. $N$.\ When each $%
c_{2^{k}-1}/c_{1}$ \ quantity (for $k=2,$\ $3$...$N)$\ is equal to\ $%
c_{2^{k}}/c_{2}$,\ this collection of relations can collectively be written
as $S_{1}$\ subset (\ref{NquibtsSubset1}), $\ $and $\mid \Psi >$\ may be
expressed as:%
\begin{equation}
\mid \Psi >=\underset{\mid \Psi (1,2,...N-1)>}{\underbrace{(\mid ++
...
+>+\frac{%
c_{4}}{c_{2}}\mid ++->+
...+
\frac{c_{2^{N}}}{c_{2}}\mid --...->)}}\otimes (c_{1}\mid
+>+c_{2}\mid ->).
\end{equation}%
\bigskip If moreover the $c_{2^{k}}/c_{2}$\ coefficients in $\mid \Psi
(1,2,...N-1)>$\ obey all the equalities expressed in Eq. (\ref%
{Subset1For(N-1)qubits}) to (\ref{Subset(N-2)For(N-1)qubits}), $\mid \Psi
(1,2,...N-1)>$ is unentangled, and $\mid \Psi >$ itself is therefore
unentangled. For instance, Eq. (\ref{Subset(N-2)For(N-1)qubits}) presently
takes the form:%
\begin{equation}
c_{2^{N-2}}c_{4\ast 2^{N-2}}=c_{2\ast 2^{N-2}} c_{3\ast 2^{N-2}}
\end{equation}%
which is subset (\ref{Subset(N-1)ForNqubits}). More generally, Eq. (\ref%
{Subset1For(N-1)qubits}) to (\ref{Subset(N-2)For(N-1)qubits}) presently take
the form of Eq. (\ref{NqubitsSubset2}) to (\ref{Subset(N-1)ForNqubits})
respectively. Any reader aiming at establishing these equalities should
appreciate that the $c_{i}$ coefficients in Eq. (\ref{Definition_ciNqubits})
are generic quantities.\ For instance, $c_{1}$ is the coefficient for $\mid
+++>$ if $N=3,$ whereas it is the coefficient for $\mid +++++>$ if $N=5.$

Conversely, if $\mid \Psi (1,2,...,N>$\ is unentangled, it can be written as:%
\begin{equation}
\underset{\text{unentangled}}{\underbrace{(\sum_{i=1}^{2^{N-1}}C_{i}\mid i>)}%
}\otimes (a_{N}\mid +>+b_{N}\mid ->)
\end{equation}%
where the $C_{i}$ coefficients obey equalities expressed through subsets (%
\ref{Subset1For(N-1)qubits}) to (\ref{Subset(N-2)For(N-1)qubits}), \ with
the $C_{i}$ instead of the $c_{i}$ coefficients. Then, the method used for $%
N=3$ and $N=4$ is again used for expressing each $c_{i}$ coefficient in
expression $\mid \Psi >=\sum_{i=1}^{2^{N}}c_{i}\mid i>$ as a function of
both a $C_{i}$ coefficient and $a_{N}$ or $b_{N}.$\ This allows us to show
that subsets (\ref{NquibtsSubset1}) to (\ref{Subset(N-1)ForNqubits}) are
obeyed. Therefore, if it is true that a ket $\mid \Psi (1,2,...N-1)>$\ is
unentangled if and only if subsets (\ref{Subset1For(N-1)qubits}) to (\ref%
{Subset(N-2)For(N-1)qubits}) are all verified, then it is true that a ket $%
\mid \Psi (1,2,...N>$ is unentangled if and only if subsets (\ref%
{NquibtsSubset1}) to (\ref{Subset(N-1)ForNqubits}) are all verified.

Considering successively $N=2,$ $3$ and $4$, it was shown above that $\mid
\Psi >$ is unentangled if and only if a collection of equalities, structured
into $(N-1)$\ subsets, is obeyed. These results and this iterative
discussion from $(N-1)$ to $N$\ $\ $finally allow us to claim that subsets (%
\ref{NquibtsSubset1}) to (\ref{Subset(N-1)ForNqubits}) do collectively
express an iff condition for a ket $\mid \Psi >$ of an $N$-qubit system to
be unentangled.

If e.g. $N=7$ (the dimension of the state space is $128$ then), it is
tedious but quite possible, through successive iterations, to get the
explicit expressions of the six subsets of equalities expressing
unentanglement. They are respectively%
\begin{eqnarray}
S_{1} &:&\ \frac{c_{1}}{c_{2}}=\frac{c_{3}}{c_{4}}=...=\frac{c_{125}}{c_{126}%
}=\frac{c_{127}}{c_{128}} \\
S_{2} &:&\text{\ \ }\frac{c_{2}}{c_{4}}=\frac{c_{6}}{c_{8}}=...=\frac{c_{126}%
}{c_{128}} \\
S_{3} &:&\ \ \frac{c_{4}}{c_{8}}=\frac{c_{12}}{c_{16}}=...=\frac{c_{124}}{%
c_{128}} \\
S_{4} &:&\ \ \frac{c_{8}}{c_{16}}=\frac{c_{24}}{c_{32}}=...=\frac{c_{120}}{%
c_{128}} \\
S_{5}:\ &&\frac{c_{16}}{c_{32}}=\frac{c_{48}}{c_{64}}=\frac{c_{80}}{c_{96}}=%
\frac{c_{112}}{c_{128}} \\
S_{6} &:&\ \ c_{32}c_{128}=c_{64}c_{96}.
\end{eqnarray}%
They have been written here in order to help the reader interpret the dots
in Eq.\ (\ref{Subset...For(N-1)qubits}) and (\ref{NquibtsSubset...}).

Our results for $N=2,$\ $3$ and $4$ suggest that there are $[2^{N}-(N+1)]$
independent equalities in Eqs. (\ref{NquibtsSubset1}) to (\ref%
{Subset(N-1)ForNqubits}). We now establish this result, again using
mathe\-matical induction.\ We first suppose that it is true that Eqs. (\ref%
{Subset1For(N-1)qubits}) to (\ref{Subset(N-2)For(N-1)qubits}), for an $(N-1)$%
-qubit system, contain $(2^{N-1}-N)$ such equalities.\ Now considering an 
$N$-qubit system, we may claim that the corresponding number of such equalities
is the sum of two quantities: $(2^{N-1}-1),$ the number of equalities
associated with $S_{1}$ (cf.\ Eq. (\ref{NquibtsSubset1})) and $(2^{N-1}-N)\ $%
\ new equalities expressing that $\mid \Psi (1,2,3,...N-1)>$ in Eq. (\ref%
{PsiproduitPsi(1.N-1)*Psi(N)}) is itself unentangled, which does lead to a
total of $[2^{N}-(N+1)]$ such equalities. This expression, which was already
known to be valid for $N=2,$ $3$ and $4$, is therefore valid for any $N\geq
2.$

\section{Discussion\label{SectionDiscussion}}

It is possible to build other sets of equalities which are obeyed if and
only if an arbitrary pure state $\mid \Psi >$ is unentangled.\ When $N=3$,
for instance, keeping the same approach, it is easy to replace (\ref%
{Subset2forN=3}) with condition $c_{1}c_{7}=c_{3}c_{5}.$\ Then, with the
same approach for $N>3,~$all even $c_{i}$ coefficients in the subsets $S_{k}$
with $k>1$ are suppressed. For $N=6,$ e.g. the $S_{5}$ subset becomes $%
c_{1}c_{49}=c_{17}c_{33}.$\ Use of even indices leads to simpler
expressions, which explains the choice made in this paper.

In \cite{DevilleA2017}, with $N=2,$ if at least one of the $c_{i}$
coefficients is equal to $0,$ it was shown that condition $%
c_{1}c_{4}=c_{2}c_{3}$ is still valid. In Section \ref{SectionFinding a
necessary and sufficient} of the present paper, it was assumed that $%
c_{i}\neq 0$ for any $i.$ When $N=3,$ if e.g. $c_{5}=0,$ then in Eq. (\ref%
{PsiArbitraireN=3}) the $(c_{5}/c_{1})\mid -+>$ term is absent and$\mid \Psi
>$ is then unentangled only if $c_{6}=0$ (cf. the presence of the ($%
c_{6}/c_{2})\mid -+>$ term in Eq. (\ref{PsiArbitraireN=3})).\ When $N>3$, if 
$c_{5}=0,$ then in all the subsets expressing unentanglement, the $c_{6}$
terms will be absent. The reason is that in Eq. (\ref%
{PsiArbitraireNArbitraire}) the (unexplicitly written) $c_{5}/c_{1}$ term of
qubits $1$ to $(N-1)$ is associated with the $c_{1}\mid +>$ state of qubit $%
N,$ and the corresponding state of qubits $1$ to $(N-1)$ associated with the 
$c_{2}\mid ->$ state of qubit $N$ has a $c_{6}/c_{2}$ coefficient. This
reasoning may also be used if more than one $c_{i}$ coefficient is equal to $%
0.$

When $N=20,$ the dimension of the state space $\mathcal{E}_{20}$ is $2^{20},$
which is roughly $10^{6}.$ An unentangled normalized state then depends upon 
$40$ real numbers only.\ In the context of Quantum Information Processing,
it is generally considered that the wealth of the quantum behaviour
originates in the existence of entanglement, but it may be important to be
able to decide whether a given pure state is entangled or not, e.g. in order
to achieve BQSS or BQPT, and finding an \textit{iff }condition is therefore
significant. The present paper has shown that, when $\mid \Psi >$ is
unentangled, there exist $[2^{N}-(N+1)]$ independent equalities between the $%
c_{i}$ coefficients, the value of which is itself roughly $10^{6}$ when $%
N=20 $.\ But it has also been found that these equalities may be classified
into only $(N-1)$ subsets, e.g.\ $19$ subsets when $N=20.$ It is hoped that
this classification should allow tractable operations in numerical
simulations or calculations.

We now come to recent papers making use of a Local Unitary (LU)
transformation. Ninety years after the building of modern Quantum Mechanics
(QM), there is a vast literature devoted to its foundations (see e.g. \cite%
{Bell1990}, \cite{Laloe2012}, \cite{Hooft2016}), while most physicists
use QM without discussing its deep content. The following lines just aim at
drawing a link between these recent papers and this literature.\ Paty \cite%
{Paty1995} has stressed that, historically, well before the 1935 EPR paper,
Einstein, at the 1927 Solvay Congress \cite{ElectronsEtPhotons1928} (p.
256), exposed his concern about what he would later on call the
incompleteness of QM. In his 1995 paper, Paty clearly and convincingly
asserts that "\textit{it is only recently, indeed, that the concept of
non-locality as a fundamental feature of quantum mechanics has been fully
appreciated, and commentators have seldom realized that this was one of
Einstein's main points", }and that\textit{\ "it is in the
Einstein-Podolsky-Rosen's paper itself that non-locality is described and
that emphasis is put on it".\ }At the same 1927 meeting, Einstein stated
that the interpretation of $\mid \Psi \mid ^{2}$ as a probability density
for a single particle (rather than for an ensemble of particles) implied,
for him, "\textit{a contradiction with the principle of relativity"}.\textit{%
\ }After 1945, a deepening of the foundations of QM partly overlapped with
the development of Classical and Quantum Information. Bell's contributions
to these foundations, from his 1966/1964 papers down to his death in 1990 
\cite{Bell1990}, especially stimulated the development of both experimental
tests and so-called quantum communications. In the context of quantum
communications, speaking of LU transformations (cf. Section \ref%
{ExistingCriteria}) is usual. An LU transformation $U=U(1)\otimes U(2)$
transforms a pure unentangled state $\mid \Psi (1>\otimes \mid \Phi (2)>$\
into the unentangled state $(U(1)$ $\mid \Psi (1>)$ $\otimes $ $(U(2)\mid
\Phi (2)>).$\ On the contrary, a simple calculation shows that e.g. the
non-LU transformation $U=e^{ias_{1x}s_{2x}}$ acting on the unentangled state 
$\mid +->,$($a$ \ is \ some dimensional real constant) transforms it into
the \textit{entangled} state \ (cf. Eq. (\ref{cns_NonIntrication2Qubits}))%
\begin{equation}
\frac{e^{ia/4}}{4}(\mid +x,+x>-\mid -x,-x>)-\frac{e^{-ia/4}}{4}(\mid
+x,-x>-\mid -x,+x>)
\end{equation}%
where e.g. $\mid +x,-x>$ means $\mid 1,+x>\otimes \mid 2,-x>$, and $\mid
i,+x>$ \ (resp.\ $\mid i,-x>)$ is the eigenket for $s_{ix}$ for the
eigenvalue $1/2$ (resp. $-1/2$), with $i=1$, $2$. LU transformations should
therefore be distinguished from entanglement-inducing transformations. The
latter transformations are faced e.g. in BPQT and BQSS, which are important
quantum information processing problems due to their applications, including
those presented hereafter.

Blind or non-blind QPT may be defined as the identification (i.e.
estimation) of a given quantum process or gate, called the direct process or
gate hereafter, which receives a \textquotedblleft source
state\textquotedblright . As discussed e.g. in{\Large \ }\cite{Branderhorst}%
, \cite{Merkel2013}, \cite{Nielsen2000}, \cite{Shukla2014}, \cite%
{Takahashi2013}, \cite{White2007}, (B)QPT is a major quantum information
processing tool, since it especially allows one to characterize the actual
behavior of quantum gates, which are the building blocks of the quantum
computers considered in Section \ref{SectionIntroduction copy(1)}. The
usual, i.e. non-blind, version of QPT requires one to know, hence to
precisely control (i.e. prepare), the specific quantum source states used as
inputs of the quantum gate to be characterized. The blind version of this
tool, i.e. BQPT, then provides an attractive extension of QPT, since it
allows one to use quantum source states whose values are unknown and
arbitrary, except that they are requested to meet some general properties.
These properties e.g. consist of unentanglement \cite{DevilleY2017IFAC},
which is one of the motivations for analyzing unentanglement in the present
paper (more details about the operation of BQPT are available e.g. in \cite%
{DevilleY2017IFAC}, \cite{DevilleY2017}).

BQSS may be seen as a quantum information processing problem where one aims at
handling the altered quantum state available at the output of a direct
quantum process / gate which typically involves undesired coupling between
its qubits, this process and its input being initially unknown. This BQSS
problem is e.g.\footnote{%
Other approaches perform BQSS directly, i.e. without first resorting to BQPT.%
} handled by (i) first identifying that direct gate with BQPT, thus using
only the output of that gate and unentanglement or other properties, (ii)
then deriving a quantum gate that performs the inverse transform of that of
the direct gate, and (iii) then feeding that \textquotedblleft inverse
gate\textquotedblright\ with the altered states available at the output of
the direct gate during final operation, so as to restore the corresponding
source, i.e. non-altered, states. One may anticipate that this approach will
be useful e.g. in situations where data are stored in a register of qubits
of a quantum computer, for subsequent use. Due to non-idealities of the
physical implementation of that register, the qubits which form it may be
coupled (e.g. when qubits are implemented as the spins of electrons which
are close to one another). As time goes on, the register state will
therefore evolve in a complicated way due to qubit coupling, thus making the
final value of that register state not directly usable in the target
application of the quantum computer. BQSS may then be used to restore the
initially stored register state, before providing it to the part of the
quantum computer which uses these nonaltered data to perform the target task
of that computer.

\section{Conclusion}

In the 2009 review article devoted to entanglement \cite{Horodecki2009}, the
Horodecki team noticed that \textit{"it appears that this new resource is
complex and difficult to detect". }Experimental and theoretical aspects were
both involved. If one focuses this comment on the idea that establishing
whether a pure state is entangled or not is a cumbersome task, the following
remarks may be made. In the present paper, devoted to an arbitrary number, $%
N,$ of distinguishable qubits, it has been shown that if a pure state of
that $N$-qubit system is developed over the $2^{N}$ basis states of the
generalized standard basis (or of some arbitrary well-defined basis) as $%
\mid \Psi >=\sum_{i}c_{i}$ $\mid i>,$ one is then led to introduce $(N-1)$
subsets of equalities, which are verified if and only if $\mid \Psi >$ is
unentangled. It should however be realized that if $N=20,$ these $19$
subsets together collect $[2^{N}-(N+1)]$ equalities, which is here roughly
equal to $2^{20},\ $i.e. approximalely $10^{6}.$ While the complexity of the
problem is reflected in the fact that the number of equalities roughly grows
as $2^{N},$ it is hoped that the necessary and sufficient condition
established in this paper, which introduced a systematic ordering within
these equalities, through a classification into $(N-1)$ subsets, may in
practice help in the manipulation of the entanglement concept.

%YDsourcedeb inclusion artiti115v5
\appendix
\section{%
%YDsourcefin inclusion artiti115v5
%YDsupprime artiti115v5 (remplace par ci-dessus):
% """
% \section{Appendix: 
% """
The von\ Neumann entropy and the establishment of the iff
condition}

The entropy concept, which did not appear in this paper yet, is briefly
considered here. The von Neumann entropy of a quantum system in a pure or
mixed state described by a density operator $\rho $ is the trace $S=-Tr(\rho
Ln\rho )$. This concept cannot directly be used in an attempt to find an 
\textit{iff} condition for the unentanglement of a pure state $\mid \Psi >$
of an $N-$qubit system, since its von\ Neumann entropy is zero for both
unentangled and entangled pure states. But this $N-$qubit system can be
viewed as a bipartite system $\Sigma ,$ composed of parts $\Sigma _{A}$ and $%
\Sigma _{B},$ and if $\Sigma $ is described by $\rho ,$ one may first
introduce reduced density operators $\rho _{A}=Tr_{B}\rho $ and $\rho
_{B}=Tr_{A}\rho $ (see e.g. \cite{PeresQT1995}). From now on,\ we focus on
the situation when $\rho =\mid \Psi ><\Psi \mid .$ Both $\Sigma _{A}$ and $%
\Sigma _{B}$ possess orthonormal basis states $\mid \varphi _{i}^{A}>$ and $%
\mid \chi _{i}^{B}>$ allowing to write any pure state $\mid \Psi >$ of $%
\Sigma $ as $\mid \Psi >=\sum_{i}\lambda _{i}\mid \varphi _{i}^{A}>\otimes
\mid \chi _{i}^{B}>$ (Schmidt decomposition), where the sum of the squares
of the real non-negative so-called Schmidt coefficients $\lambda _{i}$ is
equal to $1$ (see e.g. \cite{Nielsen2000}). Moreover, $\rho _{A}$ and $\rho
_{B}$ have the same eigenvalues, equal to $\lambda _{i}^{2}$ \cite%
{Nielsen2000}. One introduces the entropies for $\Sigma _{A}$ and $\Sigma
_{B},$\ respectively $S_{A}=-Tr_{A}(\rho _{A}Ln\rho _{A})${\LARGE \ }and%
{\LARGE \ }$S_{B}=-Tr_{B}(\rho _{B}Ln\rho _{B})${\LARGE ,} and, as a result
of both the Schmidt decomposition and the just mentioned property of the
eigenvalues of $\rho _{A}$ and $\rho _{B},$ $S_{A}=S_{B}=-\sum_{i}\lambda
_{i}^{2}Ln\lambda _{i}^{2}.$\ Then $S=0,$ while $S_{A}=S_{B}\geq 0,$ and $%
\mid \Psi >$ is unentangled if and only if $S_{A}$ $=$ $S_{B}\ $is equal to
zero. A means of establishing an $iff$ condition through the reduced entropy
concept therefore does in principle exist.\ But the fact that the reduced
entropy of a bipartite system is related to the Schmidt decomposition
immediately suggests that, if this concept is used as a tool for
establishing an $iff$ condition for the $c_{i}$ introduced in this paper,
the difficulty will be at least as great as the one found with the Schmidt
criterion, already discussed in Section \ref{ExistingCriteria}.

Let us first examine the two-qubit case: A is qubit $1$ and $B$ qubit $2$.
Then, keeping our previous notations, $\mid \Psi >=\sum_{i=1}^{4}$ $%
c_{i}\mid i>,$ one has first to express the condition $S_{A}=0$ as a
function of the $c_{i}$ coefficients,\ but this means: 1) calculating the
expression of $\rho _{A}\,,$ $2)$ calculating its eigenvalues, 3)
calculating $S_{A}$ and solving the equation $S_{A}=0.$\ The reader may
verify that a tedious calculation leads to our well-known result: 
\begin{equation}
(c_{1}c_{4}-c_{2}c_{3})=0.  \label{iffConditionForN=2}
\end{equation}%
\ The next simplest situation is $N=3$, and one may first introduce $\rho
_{3},$ the reduced entropy for qubit no. $3$, and focus on the corresponding
reduced entropy $S_{3}=$ $-Tr_{3}(\rho _{3}Ln\rho _{3}),$ which is zero 
\textit{iff }$\mid \Psi >$ is unentangled. This necessitates first to
calculate all the elements of the reduced density matrix $\rho _{3},$ each
one a complicated sum involving our $c_{i}$ coefficients, and secondly to
find an analytical expression for the eigenvalues of $\rho _{3}$. But, once
this is done, one knows that if and only if one and only one eigenvalue is
non-zero, and therefore equal to one, then the state is unentangled.
Considering the reduced entropy $S_{3},$ i.e manipulating sums of quantities
involving logarithms, is therefore unnecessary.

\end{document}